\documentclass[preprintnumbers,amsmath,amssymb,floatfix,11pt,prd,onecolumn,
superscriptaddress,nofootinbib]{revtex4}
\usepackage{latexsym}
\usepackage{epsfig}
\usepackage{amsmath}
\usepackage{epstopdf}
\usepackage{amssymb}
\usepackage{graphicx}
\usepackage{hyperref}
\usepackage{tabularx}
\usepackage{varioref}
\usepackage{placeins}
\newcolumntype{C}[1]{>{\centering\arraybackslash}p{#1}}

\begin{document}
\title{\bf KERR-NEWMAN-TAUB-NUT BLACK HOLE TUNNELLING RADIATION}
\author{Ayesha Zakria}
\email{ayesha.zakria@bcoew.edu.pk}
\affiliation{Department of Mathematics, Bilquis Post Graduate College for Women, Air University, Islamabad, Pakistan}
\author{Saba Qaiser}
\email{sabaqaiser2013@gmail.com}
\affiliation{Department of Mathematics, Bilquis Post Graduate College for Women, Air University, Islamabad, Pakistan}
\maketitle
\begin{center}
\textbf{\textbf{ABSTRACT}}
\end{center}

The tunnelling process always occur nearby the event horizon of the black hole. In this paper, we investigate the tunnelling radiation in the background of the Kerr-Newman-Taub-NUT black hole. The new coordinate system for Kerr-Newman-Taub-NUT is introduced, which helps us to form new line element. This line element helps us to show that new coordinate is exhaustively acted at event horizon. With the help of conservation of angular momentum, self gravitational effect and energy, we will show that Hawking's radiation is not exclusively thermal.

 \newpage
\section{Introduction}
\hspace{1cm}When gigantic stars fall into pieces at the end of their life cycle, then black holes are formed. Black hole is basically defined as an area in our universe, where gravitational pull is so high that nothing can passed through it \cite{1,2}. It is called as Black because any thing that enters into it can never ever come outside again. Black hole is bounded by the boundary known as event horizon. At the event horizon, the escaped velocity $V_\text{esc}$ becomes equal to speed of light i.e $V_\text{esc}=c$ where $c$ denotes the speed of light. In the 18th of century, John Michell and Pierre-Simon Laplace analysed black hole theory. In 1915, Einstein gave concept of General Relativity which helps us get more information about black holes. In 1916, first solution for black hole was established by German physicists and astronomer Karl Schwarzschild.\par
\hspace{1cm}In the study of thermal entity of our universe, black holes seem to be the most marvellous one. That is why many researchers have worked on their thermal properties. In 1974, black hole radiations was contrived by Stephen Hawking, these radiations are more probably called as thermal radiations. These radiations are formed by different process occurring near event horizon \cite{3,4}. According to Hawking, the black holes lose its energy and mass as well, that is why they are supposed to contract and eventually disappear just because of loss in mass. Hawking explained the existence of black hole radiation as particle tunnels just because of vaccum fluctations near horizon. There are two problems in Hawking's derivation for tunneling process, first is that the coordinate acted exhaustively at event horizon and other where the barrier is. After Hawking's achievement of thermal radiation, many researchers work on black hole properties and give a new idea that black holes have temperature and thermal radiations \cite{5,6,7,8,9,10,11,12,13,14,15,16}. They showed that these radiation are exclusively thermal.\par
But later on, in 2000, Maulik Parikh and Frank Wilczek gave correction to the Hawking's method of tunnelling \cite{17,18,19,20}. Parikh and Wilczek gave new idea for Hawking's radiation. They showed that these radiation are not exclusively thermal. Parik and Wilczek researched on tunnelling radiation for spherical-symmetric black hole. Tunneling radiation is found in many researches with different metrics with different concepts.\par
\hspace{1cm}Using this concept of tunneling radiation, in 2005, Zhang and Zhao extended the Parikh’s and Wilczek method, and obtained the tunneling radiation spectrum of an uncharged particle for the Kerr and Kerr-Newman black holes \cite{21,22}. They have introduced Painleve coordinate transformation which is regular at horizon of black hole, in order to eliminate the coordinate singularity. Also they have taken dragging coordinate transformation to make the event horizon and the infinite red-shift match with each other \cite{21,22}. Similarly, Hemming and Keski-Vakkuri have researched on the Hawking radiation for AdS black holes \cite{23}. Medved found Hawking radiation from a de-Sitter cosmological horizon \cite{24}. Similarly, Gu-Qiang Li have investigated on Hawking radiation of the Kerr-Sen black hole via tunnelling \cite{25}. All of these have followed Parikh’s and Wilczek. Now it is very clear that recently, Hawking radiation is followed by using Parikh’s and Wilczek tunneling method \cite{21,22,23,24,25,26,27,28}.\par
\hspace{1cm}We have reviewed tunneling radiation \cite{28} using Kerr-Newman-Kasuya \cite{29,30} black hole. In this research work, we are just going to follow Parik and Wilczek work with the help of axisymmetric and stationary black hole in the background of Kerr-Newman-Taub-NUT (KNTN) metric. We derive exact tunnelling rate of KNTN black hole. With the help of Parikh and Wilczek’s  idea of self-gravitation effect, we prove that Hawking's radiation is not exclusively thermal. This tunnelling process is observed on the particles nearby event horizon.\par
\hspace{1cm}This paper is disposed as follows. In Section \ref{Sec:C0}, we give brief introduction of KNTN metric. In Section \ref{Sec:C1}, dragging coordinate system
in the background of the KNTN black hole will be studied. With the help of dragging coordinate system,
the Painleve KNTN metric will be obtained in Section \ref{Sec:C2}. Moreover, the tunnelling process of KNTN black hole will be discussed in Section \ref{Sec:C3}. In Section \ref{Sec:C4}, Boltzmann factor will be discussed. Furthermore, inverse Hawking temperature will be discussed in \ref{Sec:C5}.
\section{Kerr-Newmen-Taub-Nut Metric}\label{Sec:C0}
\hspace{1cm}In this section, we will briefly review the metric of KNTN black hole. In the Boyer-Lindquist coordinates $(t, r, \theta,\phi)$, the KNTN black hole is entirely seted on, by the following four important  parameters: mass ($M$), charge ($Q$), spin parameter ($a$) and NUT (Newman-Unti-Tamburino) parameter ($n$). Thus KNTN metric is described by \cite{31,32,33}
\begin{eqnarray}\label{1}
ds^{2}&=&-\frac{1}{\Sigma}(\Delta-a^{2}\sin^{2}\theta)dt^{2}+\frac{2}{\Sigma}\big(\chi\Delta-a(\Sigma+a\chi)\sin^{2}\theta\big)dtd\phi+\frac{1}{\Sigma}\big((\Sigma+a\chi)^{2}\sin^{2}\theta-\chi^{2}\Delta\big)d\phi^{2}\notag\\&&
+\frac{\Sigma}{\Delta}dr^{2}+\Sigma d\theta^{2},
\end{eqnarray}
where $\Sigma$, $\Delta$ and $\chi$ are respectively defined by
\begin{eqnarray}\label{2}
\Sigma&=&r^{2}+(n+a\cos\theta)^{2}, \notag\\     \Delta &=&r^{2}-2Mr-n^{2}+a^{2}+Q^{2},\\ \chi &=&a\sin^{2}\theta-2n\cos\theta\notag.
\end{eqnarray}
The KNTN metric reduces to following metrics: Kerr-Taub-NUT for $(Q=0)$, Taub-NUT for $(a=Q=0)$, Kerr-Newman for $(n=0)$, Reissner-Nordstr\"{o}m for $(a=n=0)$, Kerr for $(n=Q=0)$ and Schwarzschild for $(a=n=Q=0)$.
\section{Dragging Coordinate System}\label{Sec:C1}
\hspace{1cm}In this section, we will study the importance of a dragging coordinate system for KNTN black hole.
First of all, we find values of event horizon and outer infinite red-shift. We can obtain the event horizon of black hole
by letting null supersurface equation $\Delta=0$ in Eq. (\ref{2})
\begin{equation}\label{3}
 r_{\text{H}}={M}+\sqrt{{M}^2+{n}^2-{a}^2-{Q}^2}.
\end{equation}
Now to obtain the outer infinite red-shift, let us consider $g_{00}=0$,
\begin{equation}\label{4}
 r^{\text{s}}={M}+\sqrt{{M}^2+{n}^2-{a}^2 cos^{2}\theta-{Q}^2}.
\end{equation}
Further more, we follow Parik's and Wilczek's work for Kerr-Newman-Taub-NUT space time. For this purpose,
start by finding coordinate system equivalent to Painleve's coordinate, we assume that new coordinate system is
exhaustively acted at the event horizon of KNTN black hole. Here one more thing to be observed, that event horizon
and outer infinite red shift do not occur at the same time i.e. both of them have different values. To over come this
 factor, we have to study dragging coordinate transformation. Suppose
\begin{equation}\label{5}
 \frac{d\phi}{dt_{\text{KNTN}}}=-\frac{g_{03}}{g_{33}}.
\end{equation}
The line element of KNTN black hole (i.e. Eq. (\ref{1})) can be rewritten in the dragging coordinate system. That is,
\begin{equation}\label{6}
 ds^2=\hat{g}_{00}dt^2_{\text{KNTN}}+\frac{\Sigma}{\Delta}dr^2+{\Sigma}d\theta^2,
\end{equation}
where
\begin{equation}
 {\hat{g}_{00}}=g_{00}-\frac{(g_{03})^{2}}{g_{33}}\notag.
\end{equation}
Using components of metric (\ref{1}), we obtain
\begin{equation}\label{7}
 {\hat{g}_{00}}=-\frac{\Sigma\Delta\sin^2\theta}{(\Sigma+a\chi)^2-\chi^2\Delta},
\end{equation}
where Eq. (\ref{6}) illustrates 3-dimensional hyperspace, in a 4-dimensional KNTN spacetime. Eqs. (\ref{6}) and (\ref{7}) show that analogous coordinate system stays coordinate singularity. Hence the coordinate transformation is detracting.
\section{Painleve Kerr-Newman-Taub Nut coordinate}\label{Sec:C2}
\hspace{1cm}It is very essential for us to take new transformation because dragging coordinate system have singularity on event horizon of KNTN black hole,
and it does not provide full information about Hawking's radiation is exclusively thermal or not. Taking another transformation in dragging coordinate system \cite{ 34,35}
\begin{equation}\label{8}
 dt_{\text{KNTN}}=dt+f(r,\theta)dr+g(r,\theta)d\theta.
\end{equation}
Here $f$ and $g$ are two functions depending on $r$ and $\theta$. As we are working on stationary black hole,
so time remains unchange. That is why $f$ and $g$ do not depend on $t$.
The integrability condition of Eq. (\ref{8}) agrees, if $f$ and $g$ satisfy the following expression
\begin{equation}\label{9}
 \frac{\partial f(r,\theta)}{\partial \theta}=\frac{\partial g(r,\theta)}{\partial r}.
\end{equation}
The equation for new time coordinate in dragging coordinate system is described by
\begin{equation}\label{10}
 t=t_{\text{KNTN}}-\int f(r,\theta)dr+g(r,\theta)d\theta.
\end{equation}
Now put Eq. (\ref{8}) in Eq. (\ref{1}) and to make spacetime flat in radial direction, we have an important result
\begin{equation}\label{11}
 g_{11}+\hat{g}_{00}f^{2}(r,\theta)=1.
\end{equation}
Consequently, Eq. (\ref{1}) can be expressed as
\begin{eqnarray}\label{2.1}
ds^{2}&=&\hat{g}_{00}dt^2+2\sqrt{{\hat{g}_{00}}\bigg(1-\frac{\Sigma}{\Delta}\bigg)}dtdr+dr^2 +\big[\hat{g}_{00}g^2(r,\theta)+\Sigma\big]d\theta^2+2\hat{g}_{00}g(r,\theta)dtd\theta\notag
\\&&+2\sqrt{{\hat{g}_{00}}\bigg(1-\frac{\Sigma}{\Delta}\bigg)}g(r,\theta)drd\theta.
\end{eqnarray}
By using new transformation (\ref{8}), we reach towards the new line element (\ref{2.1}),
which is known as Painleve Kerr-Newman-Taub Nut line element. Clearly, this coordinate is exhaustively acted
at event horizon. The Landau's condition of synchronization of coordinate clock \cite{36} is as follows
\begin{equation}\label{13}
 \frac{\partial}{\partial x^{i}}\bigg(-\frac{g_{0j}}{\hat{g}_{00}}\bigg)=\frac{\partial}{\partial x^{j}}\bigg(-\frac{g_{0i}}{\hat{g}_{00}}\bigg).
\end{equation}
Putting components of line element (\ref{2.1}) in Landau's condition of coordinate clock synchronization (\ref{13}), we obtain the following result
\begin{equation}\label{14}
 \frac{\partial f(r,\theta)}{\partial \theta}=\frac{\partial g(r,\theta)}{\partial r}.
\end{equation}
This result is equal to the result mentioned in Eq. (\ref{9}), showing that Landau's condition is satisfied.
So coordinate clock synchronization can be described by Painleve Kerr-Newman-Taub Nut line element.
Hence, the line element is flat Euclidean space in radial direction. Now obtain outer infinite red shift by substituting $\hat{g}_{00}=0$, $\Rightarrow \Sigma\Delta=0$. Since, Painleve KNTN coordinate is well behaved at event horizon, so $\Sigma\neq0$ but $\Delta=0$
\begin{equation}\label{15}
r^{\text{s}}_{i}={M}+\sqrt{{M}^2+{n}^2-{a}^2-{Q}^2}.
\end{equation}
The value of event horizon and outer infinite red-shift become same. This means that both of them occur at same time. It shows that geometrical optics limit can be used here and particle can be viewed.
\section{Tunnelling process}\label{Sec:C3}
The tunnelling process always occur nearby the event horizon of KNTN black hole.
We know that Hawking radiation are released by black hole due to the quantum effects near the horizon.
According to Hawking, a pair of particles are formed close to the horizon.
Let us consider a particle tunnelling at event horizon as an ellipsoid shell.
Also suppose that it will remain an ellipsoid shell during whole process.
In a simple way, we can say that, motion of particle along $\theta$ is equal to zero $\Rightarrow d\theta=0$.
Now putting $d\theta=0$ and $ds^{2}=0$ in Eq. (\ref{2.1}) and further simplifying it, we have found the following result
\begin{equation}\label{A16}
\bigg(\frac{dr}{dt}\bigg)^{2}+2\sqrt{{\hat{g}_{00}}\bigg(1-\frac{\Sigma}{\Delta}\bigg)}\frac{dr}{dt}+\hat{g}_{00}=0.
\end{equation}
It is clearly quadratic in {\Large $\frac{dr}{dt}$}, so we get
\begin{equation}\label{16}
\dot{r}=\frac{dr}{dt}=\frac{\sin\theta\big(\pm\Sigma-\sqrt{\Sigma(\Sigma-\Delta)}\big)}{\sqrt{(\Sigma+a\chi)^2\sin^2\theta-\chi^2\Delta}},
\end{equation}
where $\dot{r}$ represents derivative w.r.t $t$ and $\pm$ signs corresponds to the outgoing and ingoing geodesics.
\section{Boltzmann Factor}\label{Sec:C4}
\hspace{1cm}Moreover, we discuss Hawking's thermal radiation during tunnelling process. Let us assume,
that pair of effective particles casually found just inside the event horizon of KNTN black hole:
positive energy particle tunnels out while negative energy particle is ingested by black hole. In this investigation, we have considered uncharged positive particle, conservation of angular momentum and energy, and also
self-gravitation effect. Now for further discussion, let us consider energy of a particle (in an ellipsoid shell)
equals to $\omega$ and its angular momentum is equals to $\omega a$. As we are working on tunnelling process,
when particle tunnels out its mass is replaced by $(M-\omega)$ and angular momentum is replaced by
$(M-\omega)a$. Meantime, the event horizon diminishes, to event horizons corresponding to the cases pre- and post-diminishing which are the two turning points of the probable obstacle. The distance between two turning points depends on the energy of the outgoing particles and it is the width of the probable obstacle. Now new equation for event horizon is
\begin{equation}\label{17}
r^{'}_{+}={M-\omega}+\sqrt{({M-\omega})^2+{n}^2-{a}^2-{Q}^2}.
\end{equation}
The dragging angular velocity of KNTN black hole is given by
\begin{equation}\label{18}
\Omega^{'}_{\text{H}}={a}\big[2(M-\omega)^{2}-Q^{2}+2n^{2}+2(M-\omega)\sqrt{(M-\omega)^2+n^2-a^2-Q^2}\big]^{-1}.
\end{equation}
From Eq. (\ref{2.1}), note that $\phi$ is totally ignorable throughout this process. In order to overcome this
 factor, we use the action of the outgoing particle
\begin{equation}\label{19}
S=\int^{t_{f}}_{t_{i}}(L-P_{\varphi}\dot{\varphi})dt.
\end{equation}
By applying Wentzel-Kramers-Brillouin approximation, we get the relation between radiation particles and
imaginary part of the action, that is, \cite{37,38}
\begin{equation}\label{20}
\Gamma\thicksim\exp(-2\text{Im}S).
\end{equation}
The imaginary part mentioned in above equation can be calculated as
\begin{eqnarray}\label{21}
\text{Im}S&=&\text{Im}\bigg[\int^{r_{f}}_{r_{i}}P_{r}dr-\int^{\varphi_{f}}_{\varphi_{i}}P_{\varphi}d\varphi\bigg]\notag\\
&=&\text{Im}\bigg[\int^{r_{f}}_{r_{i}}\int^{P_{r}}_{0}dP^{'}_{r}dr-\int^{\varphi_{f}}_{\varphi_{i}}\int^{P_{\varphi}}_{0}dP^{'}_{\varphi}d\varphi\bigg].
\end{eqnarray}
With the help of Hamilton's equation, we get
\begin{eqnarray}
\dot{r}&=&\frac{dH}{dP_{r}}\bigg|_{(r;\varphi,P_{\varphi})}=\frac{d(M-\omega)}{dP_{r}},\label{22}\\
\dot{\varphi}&=&\frac{dH}{dP_{\varphi}}\bigg|_{(\varphi;r,P_{r})}.\label{23}
\end{eqnarray}
These equations help us to eliminate factor of momentum, where $dH_{(\varphi;r,P_{r})}=\Omega^{'}dJ$.
This shows that energy change of the black hole.
Substituting Eqs. (\ref{22} and (\ref{23}) in Eq. (\ref{21}), we get
\begin{equation}\label{24}
\text{Im}S=\text{Im}\bigg[\int^{M-\omega}_{M}\int^{r_{f}}_{r_{i}}\frac{dr}{\dot{r}}d(M-\omega^{'})-\int^{M-\omega}_{M}\int^{r_{f}}_{r_{i}}a\Omega^{'}_{\text{H}}\frac{dr}{\dot{r}}d(M-\omega^{'})\bigg].
\end{equation}
By putting value of $ \dot{r}$ in Eq. (\ref{24}), we obtain
\begin{equation}\label{26}
\text{Im}S=\text{Im}\bigg[\int^{M-\omega}_{M}\int^{r_{f}}_{r_{i}}(1-a\Omega^{'}_{\text{H}})\bigg(\frac{\sqrt{(\Sigma+a\chi)^{2}\sin^{2}\theta-\chi^{2}\Delta^{'}}}
{\sin\theta\big(\Sigma-\sqrt{\Sigma(\Sigma-\Delta^{'})}\big)}\bigg)drd(M-\omega^{'})\bigg],
\end{equation}
where
\begin{eqnarray}\label{27}
\Delta^{'}&=&r^{2}+a^{2}-n^{2}+Q^{2}-2r(M-\omega)\notag\\&=&(r-r^{'}_{+})\cdot(r-r^{'}_{-}),
\\ r_{i}&=&r_{+}=M+\sqrt{M^2+n^2-a^2-Q^2},\\ r_{f}&=&r^{'}_{+}=(M-\omega)+\sqrt{(M-\omega)^2+n^2-a^2-Q^2},
\end{eqnarray}
where $r_{i}$ and $r_{f}$ are the position of event horizon of KNTN black hole, $r_{i}$ is position before the
emanation of particles and $r_{f}$ is position after the emanation of particles. Here multiplying and dividing inner integrand by $\Sigma+\sqrt{\Sigma(\Sigma-\Delta^{'})}$, we get
\begin{eqnarray}\label{30}
\notag\text{Im}S&=&\text{Im}\bigg[\int^{M-\omega}_{M}\int^{r_{f}}_{r_{i}}(1-a\Omega^{'}_{\text{H}})\bigg(\frac{\sqrt{(\Sigma+a\chi)^{2}\sin^{2}\theta-\chi^{2}
\Delta^{'}}(\Sigma+\sqrt{\Sigma^2-\Sigma\Delta^{'}})}{\Sigma\sin\theta[(r-r^{'}_{+})(r-r^{'}_{-})]}\bigg)
\\&&\times drd(M-\omega^{'})\bigg].
\end{eqnarray}
Consider $r=r_{+}$ as a pole of first order. Then by using residue formula of complex analysis,
we can solve internal integral. We can use simple methods of integral for outermost integral, we obtain
\begin{eqnarray}\label{31}
 \text{Im}S&=&\pi\bigg[M^2-(M-\omega)^2+M\sqrt{M^2+n^2-a^2-Q^2}
-(M-\omega)\sqrt{(M-\omega)^2+n^2-a^2-Q^2}\notag\\&&-n^2\ln\bigg(\frac{(M-\omega)+\sqrt{(M-\omega)^2+n^2-a^2-Q^2}}{M+\sqrt{M^2+n^2-a^2-Q^2}}\bigg)\bigg].
 \end{eqnarray}
The rate of tunnelling is
\begin{eqnarray}\label{32}
\Gamma&\thicksim&\exp(-2\text{Im}S)\notag
\notag\\&=&\exp\bigg\{-2\pi\bigg[M^{2}-(M-\omega)^{2}+M\sqrt{M^2+n^2-a^2-Q^2}-(M-\omega)\notag
\\&&\times\sqrt{(M-\omega)^2+n^2-a^2-Q^2}-n^2\ln\bigg(\frac{(M-\omega)+\sqrt{(M-\omega)^2+n^2-a^2-Q^2}}{M+\sqrt{M^2+n^2-a^2-Q^2}}\bigg)\bigg]\bigg\}\notag
\\&=&\exp(\Delta S_{\text{BH}}),
 \end{eqnarray}
where
\begin{equation}\label{33}
 \Delta S_{\text{BH}}=S_{\text{BH}}(M-\omega)-S_{\text{BH}}(M).
 \end{equation}
It is the Bekenstein-Hawking entropy. Hence, Hawking radiation of  Kerr-Newman-Taub-Nut black hole is not exclusively thermal. The tunnelling rate for spherical symmetric black hole can be obtained by putting $a=0$ in Eq. (\ref{32}), we get
\begin{eqnarray}\label{34}
\Gamma&\thicksim&\exp(-2\text{Im}S)\notag\\&=&\exp\bigg\{-2\pi\bigg[2\omega\bigg(M-\frac{\omega}{2}\bigg)+M\sqrt{M^2+n^2-Q^2}-(M-\omega)\notag
\\&&\times\sqrt{(M-\omega)^2+n^2-Q^2}\notag
-n^2\ln\bigg(\frac{(M-\omega)+\sqrt{(M-\omega)^2+n^2-Q^2}}{M+\sqrt{M^2+n^2-Q^2}}\bigg)\bigg]\bigg\}\notag
\\&=&\exp(\Delta S_{\text{BH}}).
 \end{eqnarray}
Now expand Eq. (\ref{34}) and ignore higher terms of $\omega$, we get
\begin{eqnarray}\label{35}
\Gamma&\thicksim&\exp(-2\text{Im}S)\notag\\&=&\exp\bigg\{-2\pi\omega\bigg[(2M^2+n^2-Q^2)+2M\sqrt{M^2+n^2-Q^2}\bigg]
\notag\\&&\times(M^2+n^2-Q^2)^{-1/2}\bigg\}+2\pi\bigg\{[M+\sqrt{M^2+n^2-Q^2}]^{-n^2}\notag
\\&&-{\bigg[\frac{(M-\omega)\sqrt{M^2+n^2-Q^2}+(M^2+n^2-Q^2)-\omega{M}}{\sqrt{M^2+n^2-Q^2}}\bigg]^{-n^2}}\bigg\}.
\end{eqnarray}
Here Eq. (\ref{35})) is the Boltzmann factor. Therefore, Hawking radiation of spherical symmetric KNTN black hole contracts to Boltzmann factor at a certain
amount of temperature.
\begin{figure}[h!]
\centering
\includegraphics[width=8cm, height=8cm]{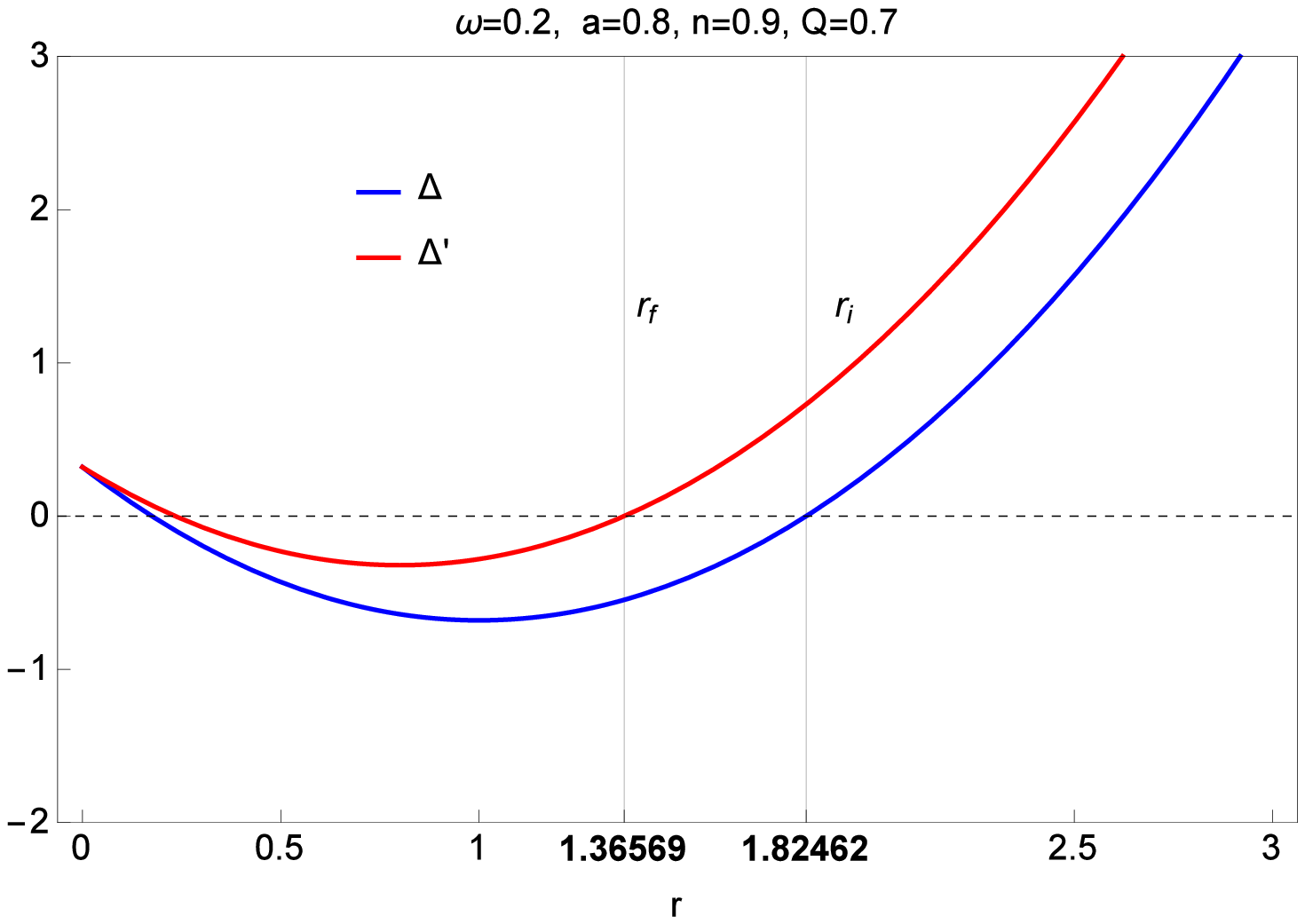}
\includegraphics[width=8cm, height=8cm]{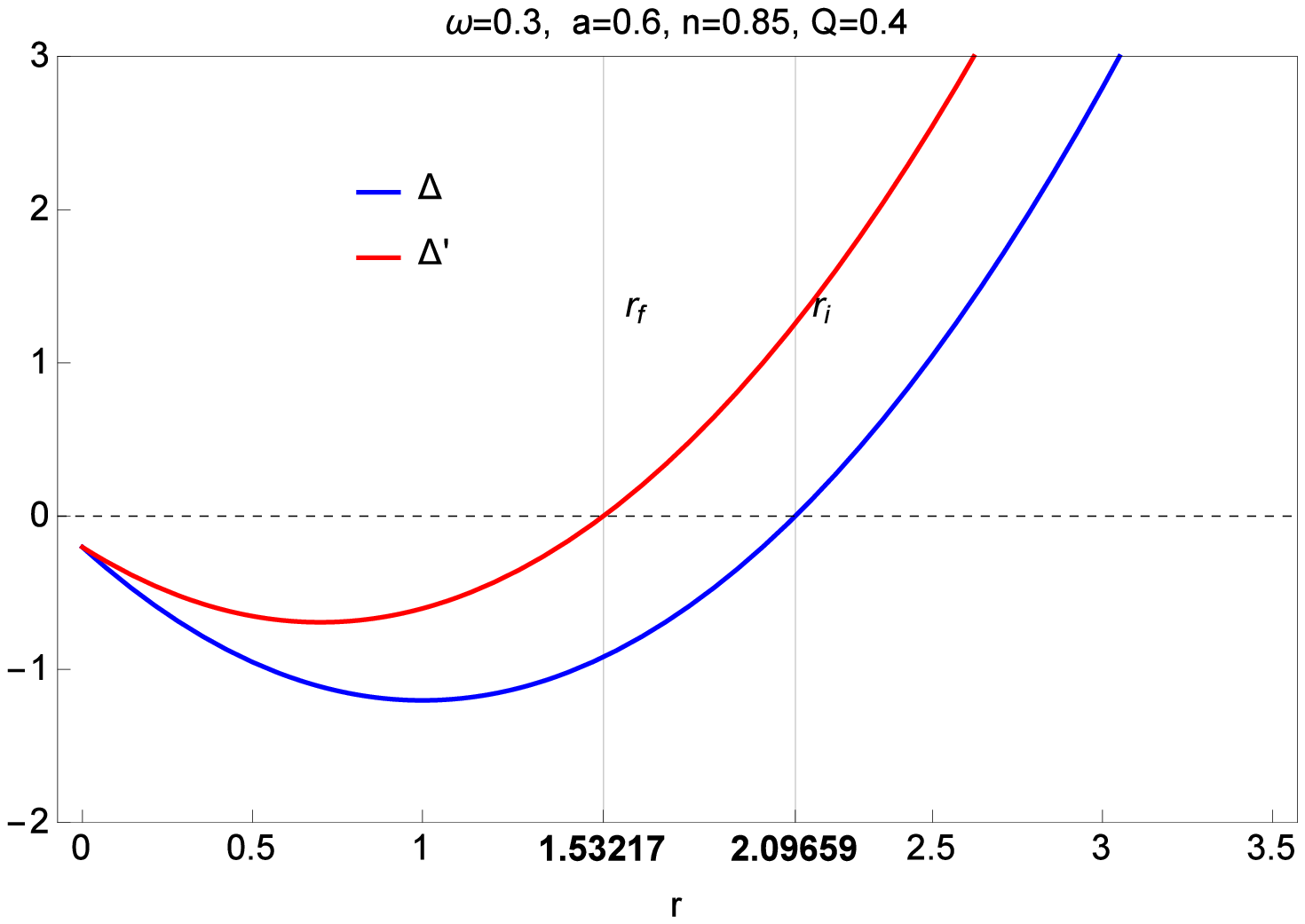}
\includegraphics[width=8cm, height=8cm]{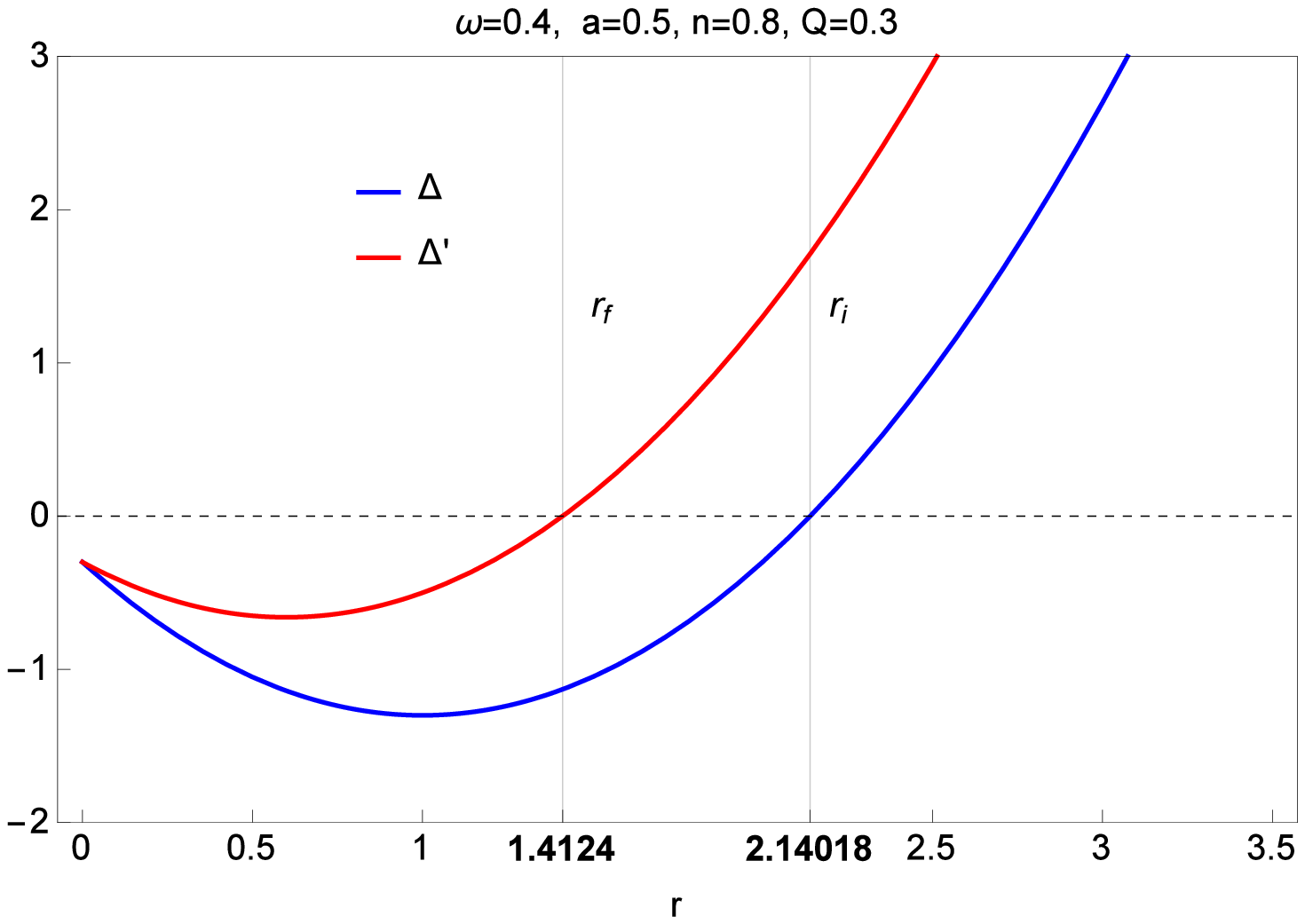}
\includegraphics[width=8cm, height=8cm]{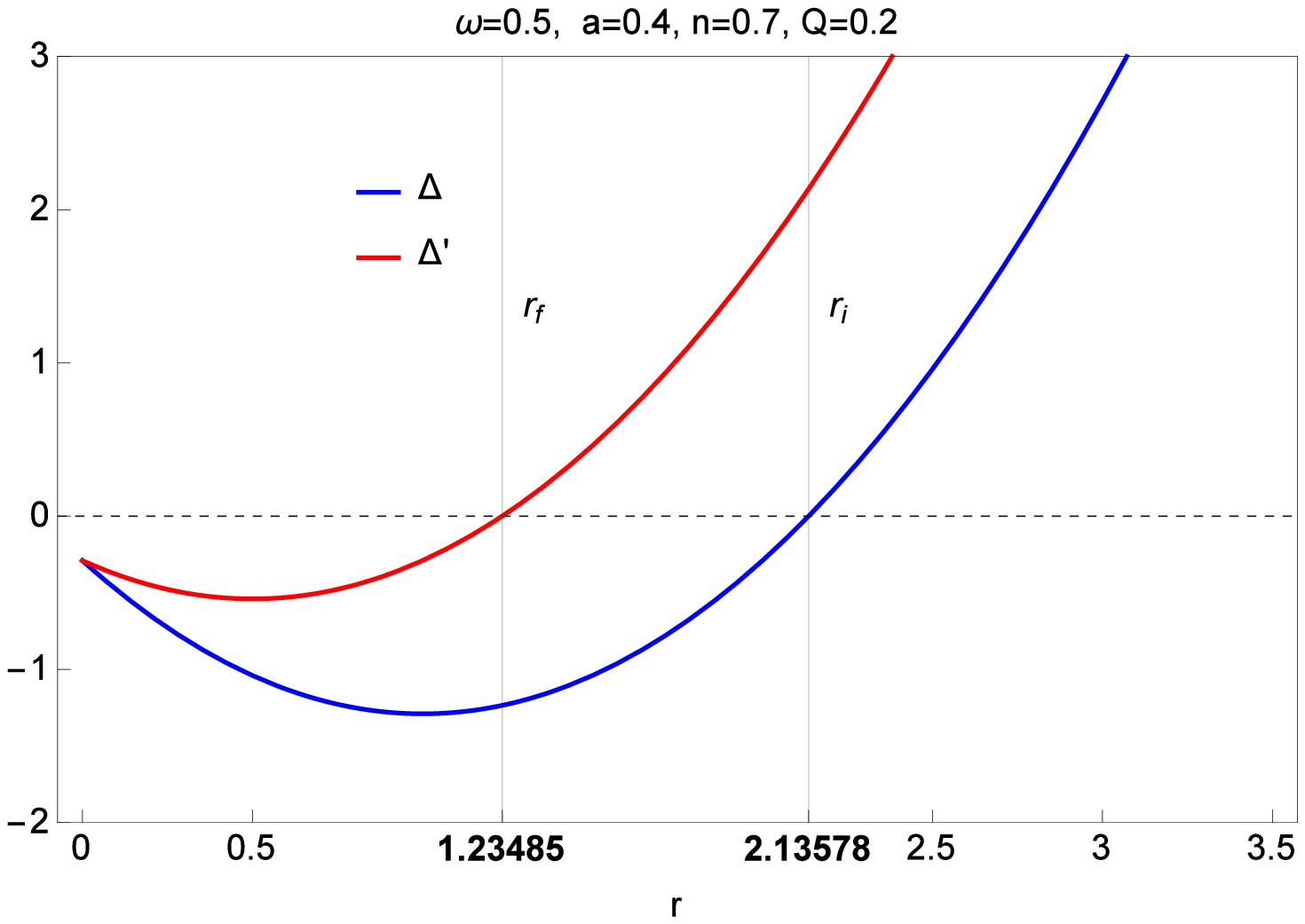}
\includegraphics[width=8cm, height=8cm]{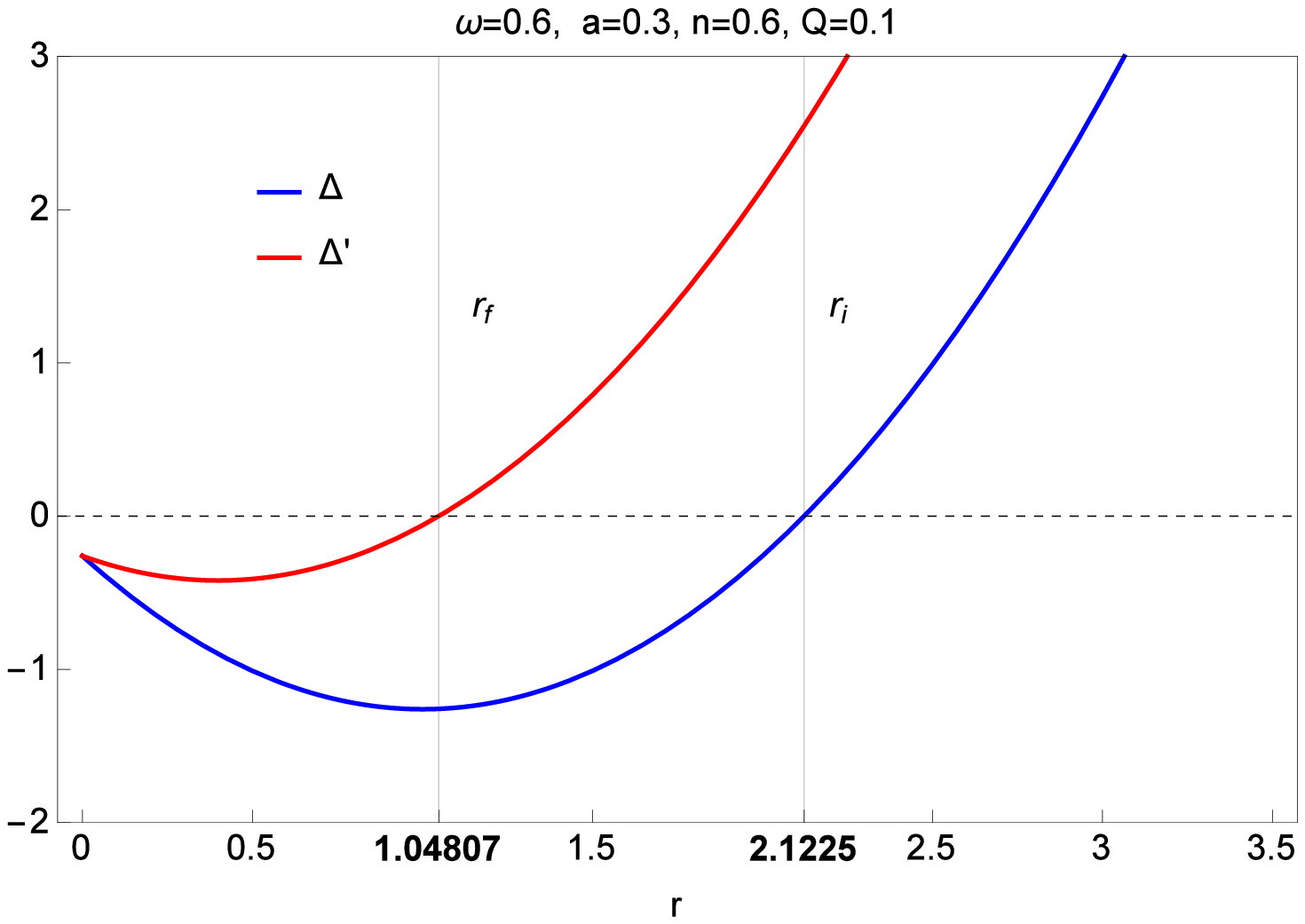}
\caption{The functions $\Delta(r)$ (blue curve) and $\Delta^{'}(r)$ (red curve) for different values of $\omega$, $a$, $n$ and $Q$. We set $M=1$. Vertical lines recognize position of event horizons before and after emanation of particles.}\label{f1}
\end{figure}
\begin{figure}[h!]
\centering
\includegraphics[width=8cm, height=8cm]{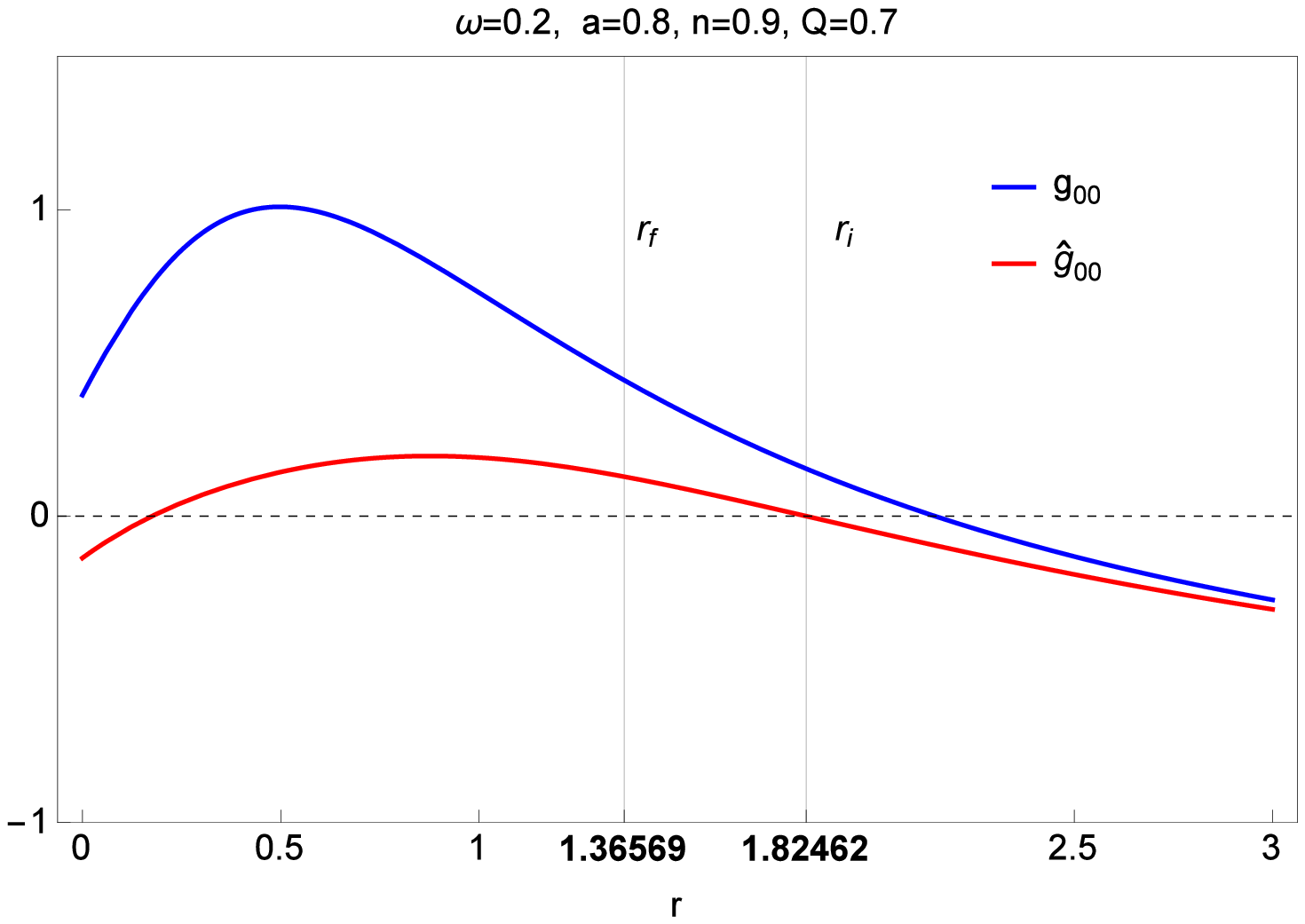}
\includegraphics[width=8cm, height=8cm]{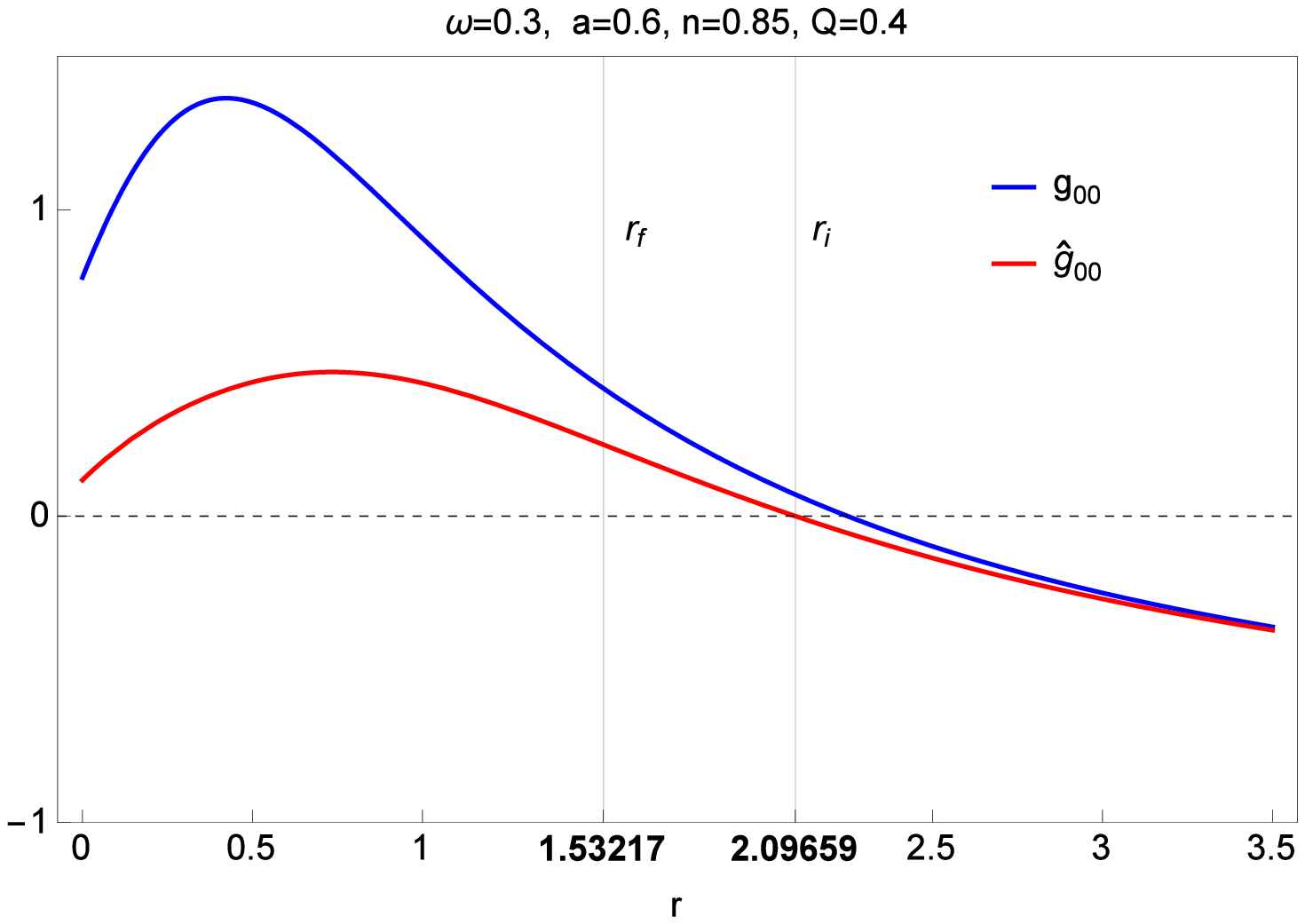}
\includegraphics[width=8cm, height=8cm]{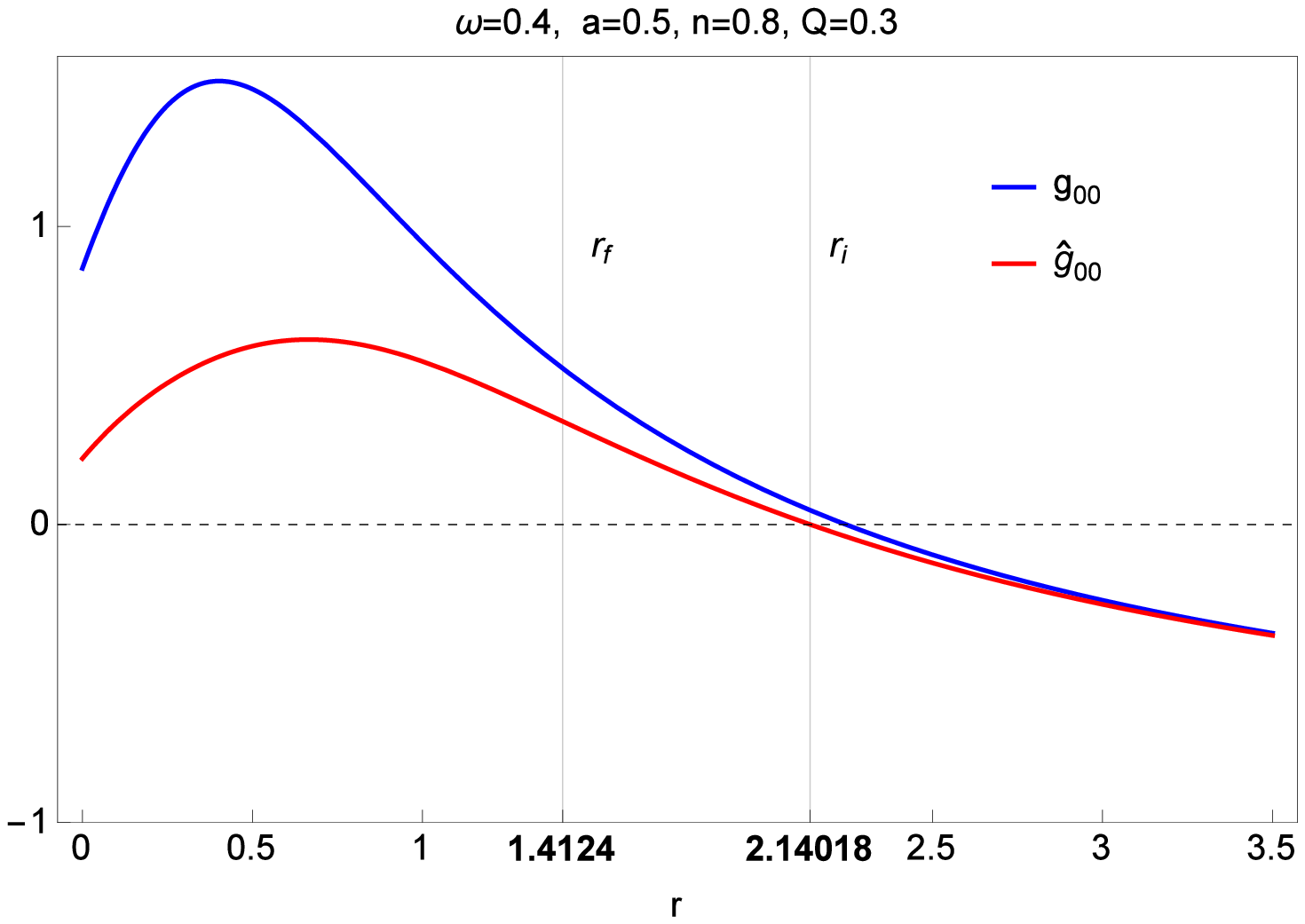}
\includegraphics[width=8cm, height=8cm]{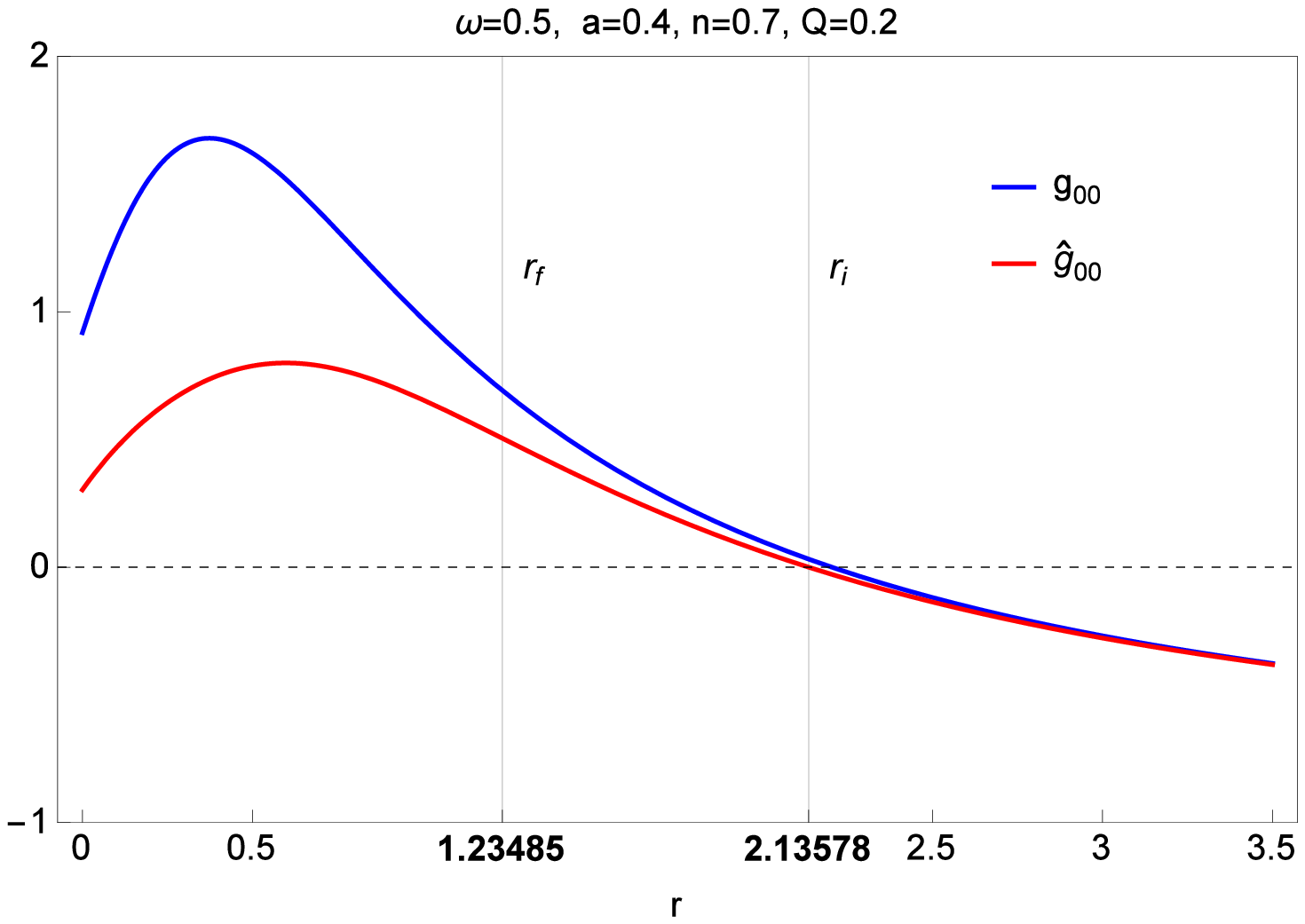}
\includegraphics[width=8cm, height=8cm]{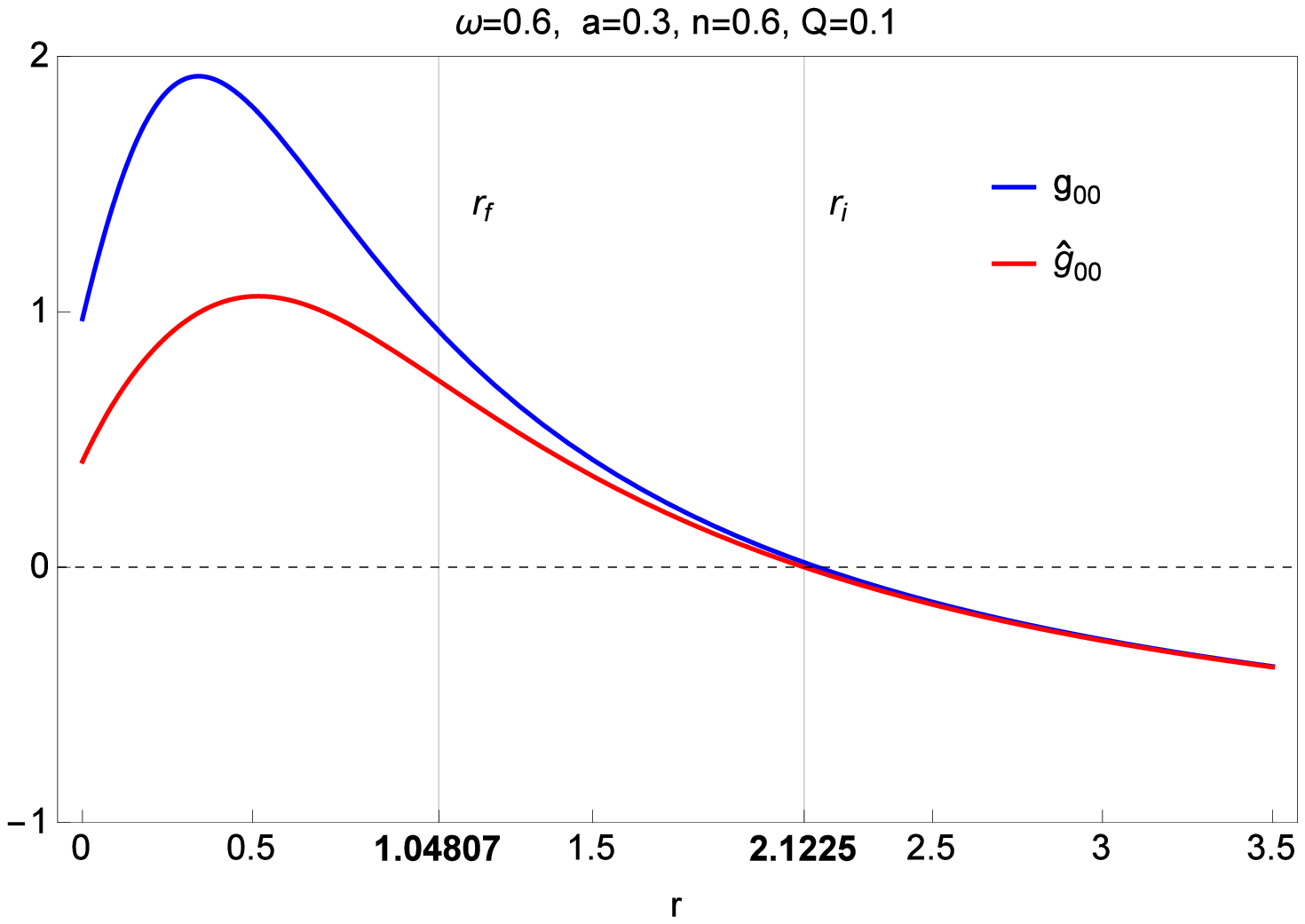}
\caption{The functions $g_{00}$ (blue curve) and $\hat{g}_{00}$ (red curve) for different values of $\omega$, $a$, $n$ and $Q$ in the equatorial plane. We set $M=1$. Vertical lines recognize position of event horizon before and after emanation of particles.}\label{f2}
\end{figure}
\begin{figure}[h!]
\centering
\includegraphics[width=8cm, height=8cm]{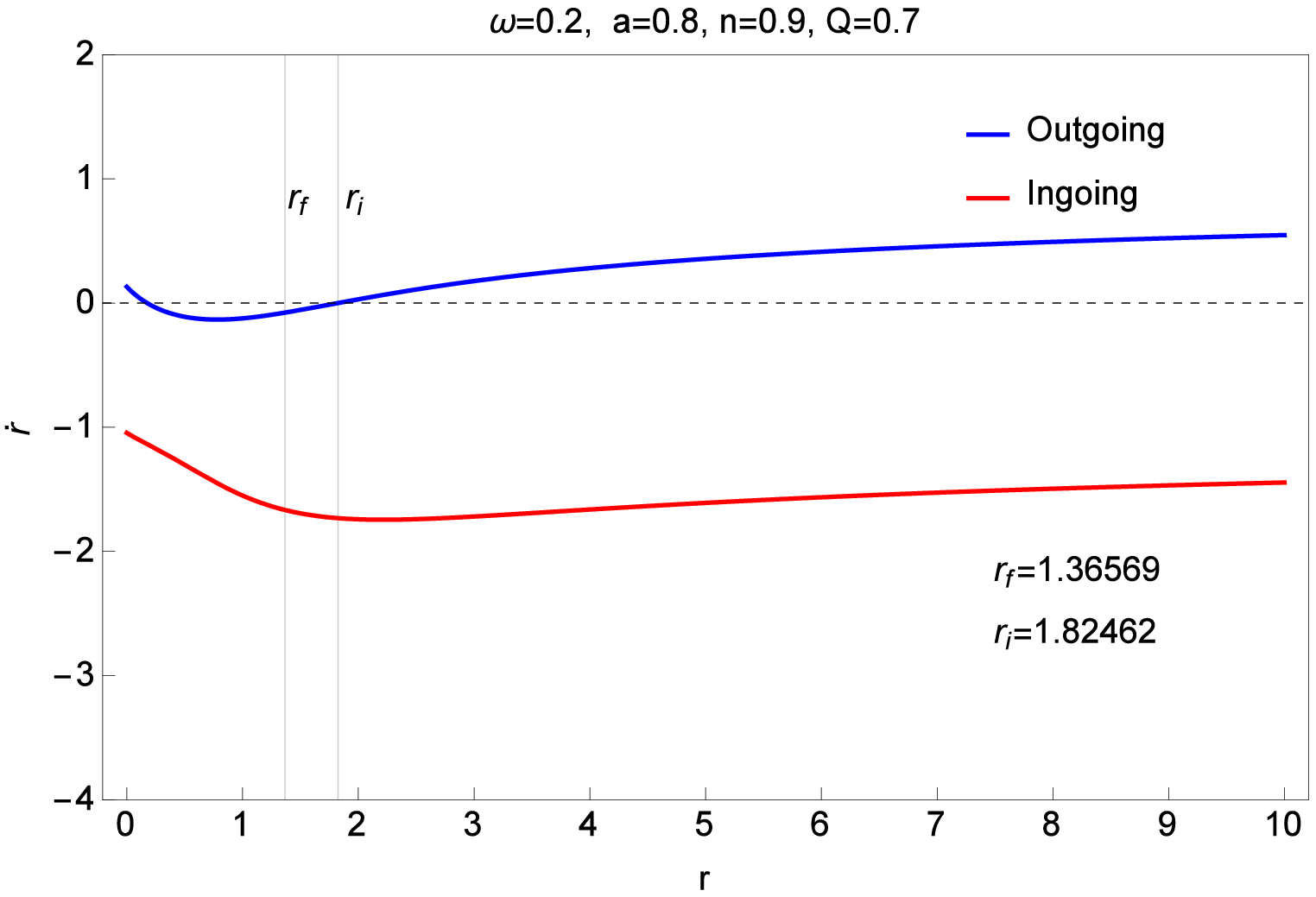}
\includegraphics[width=8cm, height=8cm]{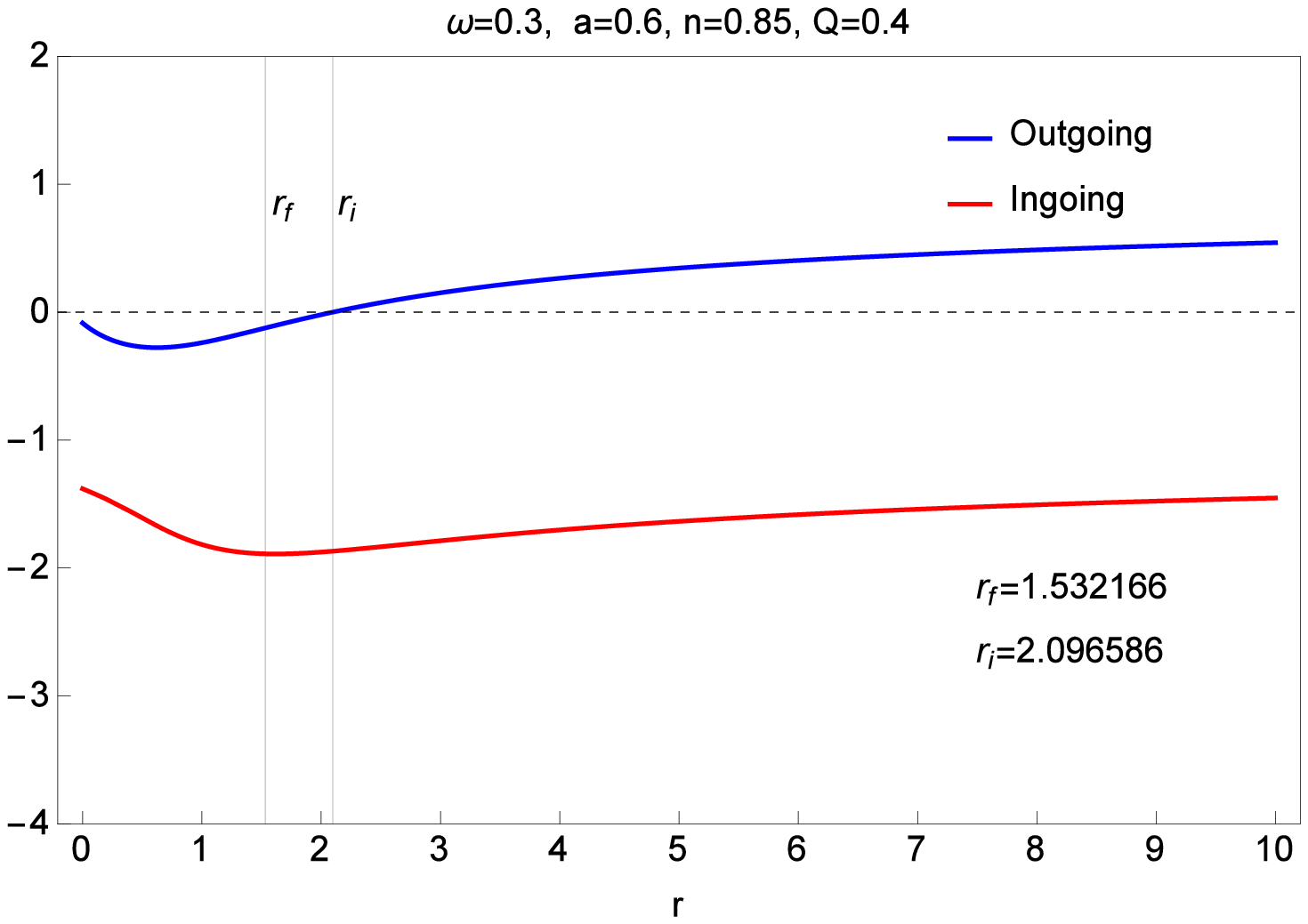}
\includegraphics[width=8cm, height=8cm]{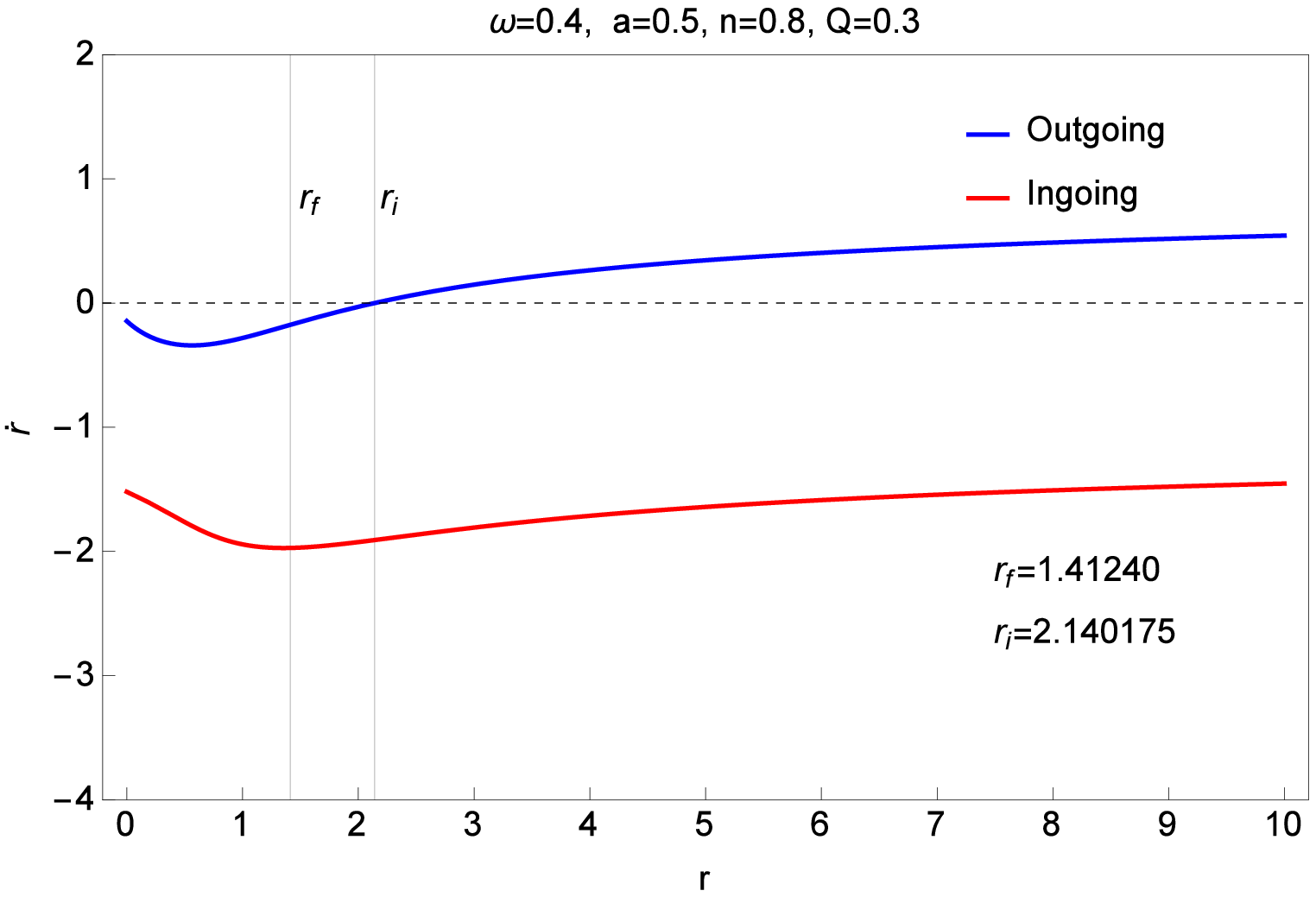}
\includegraphics[width=8cm, height=8cm]{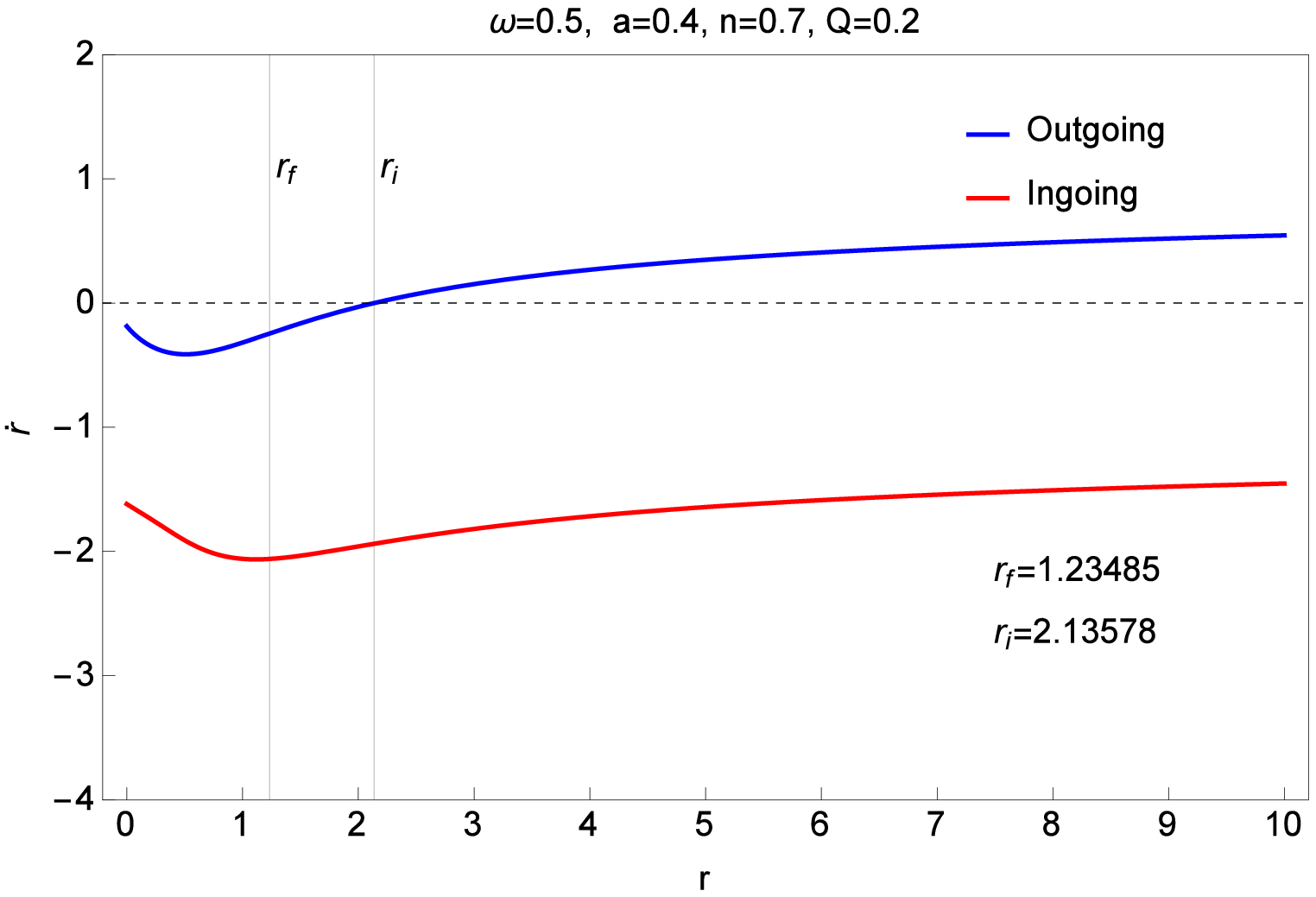}
\includegraphics[width=8cm, height=8cm]{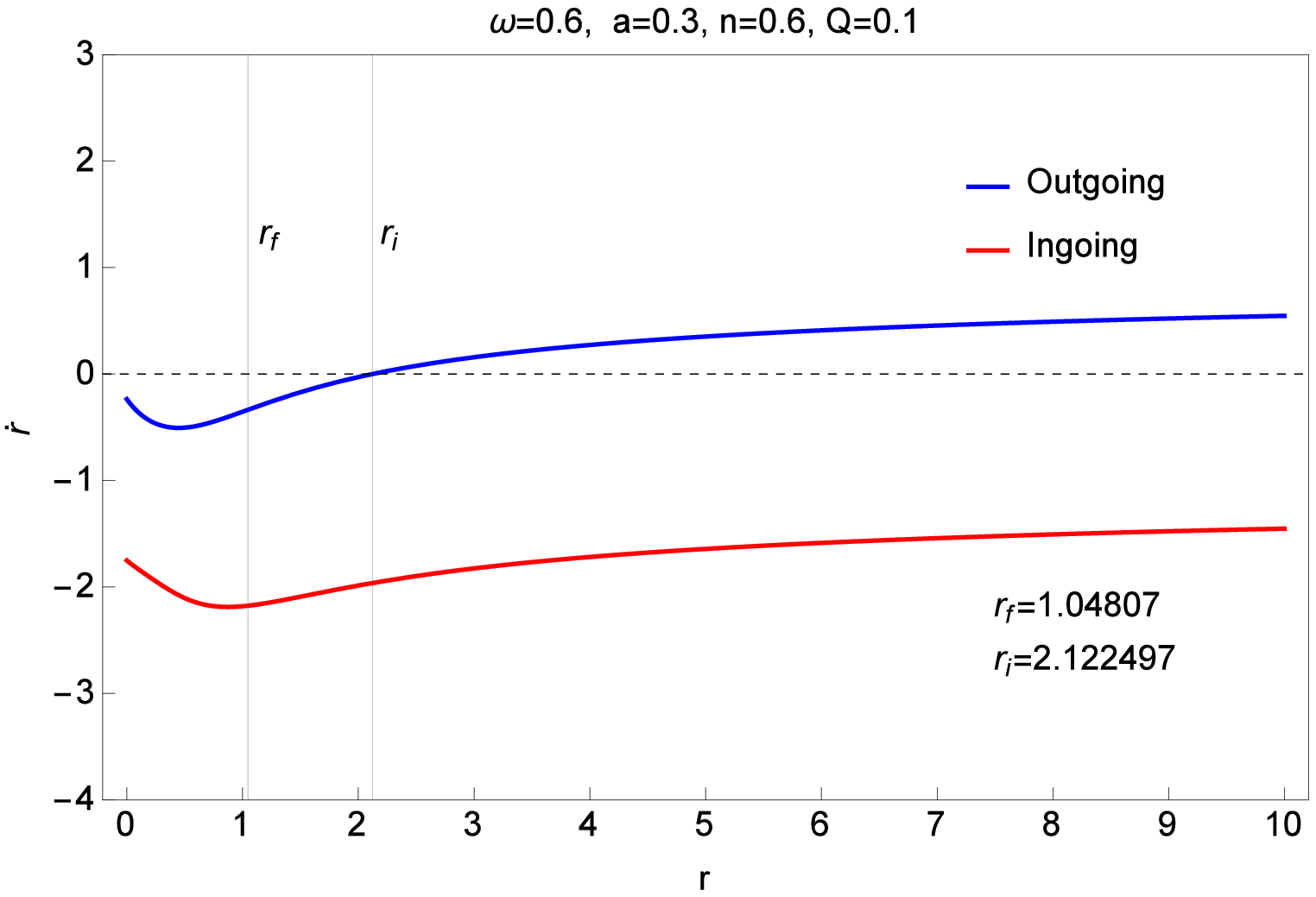}
\caption{The outgoing (blue curve) and ingoing (red curve) geodesics for different values of $\omega$, $a$, $n$ and $Q$ in the equatorial plane. We set $M=1$. Vertical lines recognize position of event horizon before and after emanation of particles.}\label{f3}
\end{figure}

We plot the functions $\Delta(r)$ (blue curve) and $\Delta^{'}(r)$ (red curve) for different values of $\omega$, $a$, $n$ and $Q$ in Figure (\ref{f1}). The functions $g_{00}$ (blue curve) and $\hat{g}_{00}$ (red curve) are shown in Figure (\ref{f2}) for $\theta=$ {\Large $\frac{\pi}{2}$}. The outgoing (blue curve) and ingoing (red curve) geodesics represent in Figure (\ref{f3}) for $\theta=$ {\Large $\frac{\pi}{2}$}. We set $M=1$. Vertical lines recognize position of event horizons before and after emanation of particles. We choose different values of $\omega$, $a$, $n$ and $Q$: (I) $\omega=0.2,~a=0.8,~n=0.9,~Q=0.7$, (II) $\omega=0.3,~a=0.6,~n=0.85,~Q=0.4$, (III) $\omega=0.4,~a=0.5,~n=0.8,~Q=0.3$, (IV) $\omega=0.5,~a=0.4,~n=0.7,~Q=0.2$ and (V) $\omega=0.6,~a=0.3,~n=0.6,~Q=0.1$.
\section{Inverse Hawking Temperature}\label{Sec:C5}
Now inverse Hawking temperature is
\begin{equation}\label{36}
T_{\text{H}}=\frac{\kappa}{2\pi},
\end{equation}
where $\kappa$ is surface density and it is equal to
\begin{equation}\label{37}
\kappa=\frac{1}{2}\frac{df}{dr}\bigg|_{r=r_{+}}=\frac{r_{+}-r_{-}}{2(r_{+}^{2}+a^{2})},
\end{equation}
where $f(r)=g_{00}|_{\theta=0}$. Thus, we have
\begin{equation}\label{38}
T_{\text{H}}=\frac{\sqrt{M^2+n^2-Q^2}}{2\pi\big(M+\sqrt{M^2+n^2-Q^2}\big)^{2}}.
\end{equation}
\section{conclusion}
\hspace{1cm}On the basis of our research, we have concluded that the tunnelling process always occur nearby the event horizon of KNTN black hole. We have found exact rate for particles, which are tunnel beyond the event horizon of KNTN black hole. In this study, we have used different methods to find the thermal radiation of black hole. With the help of conservation of angular momentum, self gravitational effect and energy, we have showed that Hawking's radiation is not exclusively thermal. The event horizon diminishes, to event horizons corresponding to the cases pre- and post-diminishing which are the two turning points of the probable obstacle. The distance between two turning points depends on the energy of the outgoing particles and it is the width of the probable obstacle.

\end{document}